\documentstyle[11pt,epsf]{article}

\newlength{\mytopmargin}
\newlength{\myleftmargin}
\def\zz{\relax\hbox{\small \sf Z\kern-.4em Z}}
\newcommand{\mbf}[1]{\mbox{\boldmath$#1$}}
\newcommand{\ml}{\langle }
\newcommand{\mg}{\rangle }

\newtheorem{lemma}{Lemma}[section]
\newtheorem{prop}[lemma]{Proposition}
\renewcommand{\theequation}{\thesection.\arabic{equation}}

\begin{document}

\vspace{1cm}
\noindent
\begin{center}{ \Large \bf
Classical skew orthogonal polynomials and random matrices}
\end{center}
\vspace{5mm}

\begin{center}
M.~Adler${}^{\rm a}$, P.J.~Forrester${}^{\rm b}$, 
T.~Nagao${}^{\rm c}$ and P.~van Moerbeke${}^{\rm d}$
\end{center}

\vspace{.2cm}
\noindent
${}^{\rm a}$ Department of Mathematics, Brandeis University, Waltham, Mass
02454, USA

\noindent
${}^{\rm b}$ Department of Mathematics and Statistics, University of
Melbourne, Parkville 3052, Australia

\noindent
${}^{\rm c}$ Department of Physics, Graduate School of Science, 
Osaka University, 
Toyonaka, Osaka 560-0043, Japan

\noindent
${}^{\rm d}$ Department of Mathematics, Universit\'e de Louvain-la-Neuve,
Belgium and  Brandeis University, Waltham, Mass 02454, USA

\vspace{.2cm}
{\small
\begin{quote}
Skew orthogonal polynomials arise in the calculation of the $n$-point
distribution function for the eigenvalues of
ensembles of random matrices with orthogonal
or symplectic symmetry. In particular, the distribution functions are
completely determined by a certain sum involving the skew orthogonal
polynomials. In the cases that the eigenvalue probability density function
involves a classical weight function, explicit formulas for the skew
orthogonal polynomials are given in terms of related orthogonal
polynomials, and the structure is used to give a closed form expression
for the sum. This theory treates all classical cases on an equal footing,
giving formulas applicable at once to the Hermite, Laguerre and
Jacobi cases.
\end{quote} 
}

\vspace{.5cm}
\noindent
\section{Introduction}
\setcounter{equation}{0}
The classical polynomials play an essential role in calculating statistical
properties of the eigenvalues of certain classes of random matrices
with normally distributed elements. As a concrete example, consider a
random $N \times N$ Hermitian matrix $X$ in which the diagonal elements
(which must be real) and the upper triangular elements are independently
chosen with normal distributions N$[0,1/\sqrt{2}]$ and
${\rm N}[0,1/2] + i {\rm N}[0,1/2]$ respectively. Then the corresponding
eigenvalue probability density function (p.d.f.) is proportional to
\begin{equation}\label{1.1}
\prod_{j=1}^N w_2(x_j) \prod_{1 \le j < k \le N}
(x_k - x_j)^2, \quad w_2(x) = e^{-x^2}.
\end{equation}
Note that this involves the classical weight function $e^{-x^2}$.

The monic orthogonal polynomials corresponding to the weight function
$w_2(x)$, $\{p_j(x)\}$ say,
 are introduced by writing the product of differences in
(\ref{1.1}) as the square of the Vandermonde determinant, and adding
together appropriate multiples of the columns. This shows that
(\ref{1.1}) is equal to
\begin{equation}\label{1.2}
\prod_{j=1}^N w_2(x_j) \Big ( \det \Big [ p_{j-1}(x_k) \Big ]_{j,k=1,\dots,N}
 \Big )^2.
\end{equation}
This form of the p.d.f.~provides the starting point for the calculation
of the $n$-point distribution functions
(see Section 3), which in turn are used in
calculating spacing distributions (see e.g.~\cite{Me91}).

The p.d.f.~(\ref{1.1}) with the classical weight function
$w_2(x) = x^a e^{-x}$, $(x > 0$), and 
$w_2(x) = x^a (1 - x)^b$, $(0 < x < 1)$, also occurs in random
matrix problems (see e.g.~\cite{Mu82}). Thus if $X$ is a rectangular
$n \times m$, $(n \ge m)$, matrix with complex elements independently
distributed according to the normal distribution
${\rm N}[0,1/\sqrt{2}] + i{\rm N}[0,1/\sqrt{2}]$, then the eigenvalues of
$X^\dagger X$ have the distribution (\ref{1.1}) with
$w_2(x) = x^{n-m} e^{-x}$ and $N = m$. Similarly, if $X_1$ and $X_2$
are rectangular matrices of dimensions $n_1 \times m$ and
$n_2 \times m$, then the eigenvalues of $X_1^\dagger X_1
(1 + X_2^\dagger X_2)^{-1}$ have the distribution (\ref{1.1}) with
$w_2(x) = x^{n_1 - m}(1 - x)^{n_2 - m}$ and $N = m$.

If the random matrices $X$ which lead to the eigenvalue p.d.f~(\ref{1.1})
with the classical weight functions noted above are specified to have
real elements, or real quaternion elements represented by $2 \times 2$
blocks of the form
$$
\left [ \begin{array}{cc} z_{jk} & w_{jk} \\
- \bar{w}_{jk} & \bar{z}_{jk} \end{array} \right ]
$$
(in which case the eigenvalues are doubly degenerate), the corresponding
eigenvalue p.d.f.~is then proportional to 
\begin{equation}\label{1.3}
\prod_{j=1}^N w_\beta (x_j) \,
\prod_{1 \le j < k \le N} |x_k - x_j |^\beta.
\end{equation}
Here $\beta = 1$ when the elements are real, while $\beta = 4$ when the
elements are real quaternion, and the weight function is again 
of the same classical form as in the complex case (\ref{1.1}) (up to
scaling of the eigenvalues $x_j$). 
We remark that ensembles of such matrices have orthogonal $(\beta = 1)$
and symplectic $(\beta = 4)$ symmetry.
Instead of writing the product of
differences in (\ref{1.3}) in terms of an ordinary determinant involving
skew orthogonal polynomials, it has been shown by Mahoux and
Mehta \cite{MM91} (see also \cite{NW91,FP95,TW98})
that for the purposes of calculating the $n$-point
distribution one should introduce a quaternion determinant involving
skew orthogonal polynomials. 
The definition of a quaternion determinant can be found for example in
\cite{Dy70}; we remark here that for the $2N \times 2N$ matrices $Q$
occuring in such a calculation (which technically are self dual) the
quaternion determinant, to be denoted Tdet, is related to the usual
Pfaffian via the formula
\begin{equation}\label{1.4}
{\rm Tdet} \, Q = {\rm Pf} (Z Q), \qquad
Z := 1_N \otimes \left ( \begin{array}{cc} 0 & - 1 \\
1 & 0 \end{array} \right ).
\end{equation}

Our interest is in the calculation of the classical skew orthogonal
polynomials in terms of classical orthogonal polynomials, and the
application of these formulas in the evaluation of the matrix
elements in the Tdet expression for the $n$-point distribution.
We begin in Section 2 by revising the relationship between the
p.d.f.~(\ref{1.3}) for $\beta = 1$ and 4 and skew orthogonal polynomials.
Also revised is the corresponding expression for the $n$-point
distribution functions in terms of the  skew orthogonal polynomials.
Next we present the recent theory of Adler and van Moerbeke
\cite{AV99} which identifies the family of orthogonal polynomials
naturally related to a particular family of skew orthogonals, and
furthermore relates the skew orthogonal polynomials at $\beta = 1$
to those at $\beta = 4$.

In Section 3 this theory is used to calculate the classical skew orthogonal
polynomials in terms of their natural basis of orthogonal polynomials.
The results are shown to be consistent with expressions obtained in
the earlier studies of Nagao and Wadati \cite{NW91}.
Application of these expansion formulas is made in Section 4, in which
the fundamental series involving the skew orthogonal polynomials
occuring in the expression of the $n$-point distribution is summed
for all the classical ensembles. In the Hermite case this reclaims
known results. For the Laguerre ensemble our summation formula differs
from that obtained in recent studies of Widom \cite{Wi98} and
Forrester et al.~\cite{FNH98} --- the reconciliation of the two seemingly
different forms is made in the Appendix. The remaining classical case is
the Jacobi ensemble. Here our summation formulas constitute new results.

\section{Revision}
\setcounter{equation}{0}
\subsection{Skew orthogonal polynomials and the $n$-point distribution}\label{s2.1}
The $n$-point distribution $\rho_{(n)}$ with respect to the p.d.f.~(\ref{1.3})
is defined as a ratio of multiple integrals according to
\begin{equation}\label{2.1}
\rho_{(n)}(x_1,\dots,x_n) = \prod_{j=1}^n w_\beta (x_j)
{\prod_{l=n+1}^N \int_{-\infty}^\infty dx_l \, w_\beta(x_l)
\prod_{1 \le j < k \le N} |x_k - x_j |^\beta \over
\prod_{l=1}^N  \int_{-\infty}^\infty dx_l \, w_\beta(x_l)
\prod_{1 \le j < k \le N} |x_k - x_j |^\beta}.
\end{equation}
Generalizing the pioneering work of Dyson \cite{Dy70}, it has been shown
by Mahoux and Mehta \cite{MM91} that for $\beta = 1$ and $\beta = 4$
$\rho_{(n)}$ can be evaluated as an $n \times n$ quaternion determinant
involving certain skew orthogonal polynomials.

In preparation for presenting these results, we recall that an inner
product $\ml f, g \mg$ is referred to as skew if
$$
\ml f, g \mg = - \ml g, f \mg
$$
while a family of (monic) polynomials $\{ q_n(x) \}_{n=0,1,2,\dots}$
are said to be skew orthogonal if
\begin{equation}\label{2.2'}
\ml q_{2m}, q_{2n+1} \mg = - \ml q_{2n+1}, q_{2m} \mg = r_m \delta_{m,n},
\qquad \ml q_{2m}, q_{2n} \mg = \ml q_{2m+1}, q_{2n+1} \mg = 0.
\end{equation}
Equivalently, for any given $N$ we require
\begin{equation}\label{2.2}
[ \ml q_j, q_k \mg ]_{j,k=0,\dots,2N-1} = { R}, \quad
 R := \left [
\begin{array}{ccccccc} 0 & r_1 & & & & & \\
-r_1 & 0 & & & & & \\
 & & 0 & r_2 & & & \\
& & -r_2 & 0 & & & \\
 & & & & \ddots & & \\
 & & & & & 0 & r_N \\
 & & & & & -r_N & 0
\end{array}
\right ]
\end{equation}
We remark that the relations (\ref{2.2'}) are unchanged under the 
replacement
\begin{equation}\label{2.re}
q_{2m+1}(x) \mapsto q_{2m+1}(x) + \alpha_{2m} q_{2m}(x)
\end{equation}
for arbitrary $\alpha_{2m}$, so the skew orthogonal transformations are 
non-unique up to this mapping.

The specific skew inner products of relevance to the calculation of
(\ref{2.2}) are
\begin{equation}\label{2.3}
\ml f, g \mg_1 := {1 \over 2} \int_{-\infty}^\infty dx \,
e^{-V(x)} \int_{-\infty}^\infty dy \, e^{-V(y)}
{\rm sgn} (y-x) f(x) g(y)
\end{equation}
and
\begin{equation}\label{2.4}
\ml f, g \mg_4 := {1 \over 2}  \int_{-\infty}^\infty dx \,
e^{-2V(x)} \Big ( f(x) g'(x) - f'(x) g(x) \Big )
\end{equation}
for $\beta = 1$ and $\beta = 4$ respectively, where in (\ref{2.2})
we have chosen
\begin{equation}\label{2.6'}
w_1(x) = e^{-V(x)} \qquad w_4(x) = e^{-2 V(x)}.
\end{equation}
We denote the corresponding monic skew orthogonal polynomials by
$\{ q_j^{(1)} \}_{j=0,1,\dots}$ and $\{ q_j^{(4)} \}_{j=0,1,\dots}$
and apply similar superscripts to the matrix $ R$ and its elements.

With these preliminaries, the quaternion determinant formulas for
(\ref{2.1}) can now be presented.

\subsection*{$\beta = 1$, $N$ even}
For $\beta = 1$ the result depends on the parity of $N$. Let $N$ be
even and set
\begin{eqnarray}
\Phi_k(x) & := &
{1 \over 2} \int_{-\infty}^{\infty}{\rm sgn}(x-y) q_k^{(1)}(y)
e^{-V(y)} \, dy, \label{2.Phi}\\
 f_1(x,y) & = &
 \left[
\begin{array}{ll}
S_1(x,y) & I_1(x,y) \\
D_1(x,y) & S_1(y,x)
\end{array} \right] \nonumber
\end{eqnarray}
with
\begin{eqnarray}
S_1(x,y) & = & \sum_{k=0}^{N/2-1} {e^{-V(y)} \over r_k^{(1)}} 
\Big ( \Phi_{2k}(x)q_{2k
+1}^{(1)}(y) -
\Phi_{2k+1}(x)q_{2k}^{(1)}(y) \Big ) \nonumber\\
D_1(x,y) & = & {\partial \over \partial x} S_1(x,y) \nonumber \\
I_1(x,y) & = & {1 \over 2} \int_{-\infty}^{\infty}S_1(x,z) {\rm sgn} (z-y) dz - {1 \over
2} {\rm
sgn}(x-y) \label{2.8a}
\end{eqnarray}
Then with $w_1(x) = e^{-V(x)}$ in (\ref{2.1}) we have \cite{MM91}
\begin{equation}\label{2.7}
\rho_{(n)}(x_1,\dots,x_n) = {\rm qdet}[ f_1(x_j,x_k)]_{j,k=1,\dots,n}.
\end{equation}

\subsection*{$\beta = 1$, $N$ odd}
Let $\{q_n^{(1)} \}_{n=0,1,\dots}$ and $\{r_n^{(1)} \}_{n=0, \dots,
(N-1)/2 - 1}$ be as in the $N$ even case, and define
\begin{eqnarray}
r_{(N-1)/2}^{(1)} &:=& {1 \over 2} \int_{-\infty}^\infty dx \,
e^{-V(x)} q_{N-1}^{(1)}(x) \nonumber \\
\hat{q}_n^{(1)}(x) &:=& q_n^{(1)}(x) - {1 \over 2 r_{(N-1)/2}^{(1)}}
\Big ( \int_{-\infty}^\infty dx' \, e^{-V(x')} q_n^{(1)}(x')
 \Big ) q_{N-1}^{(1)}
(x) \quad (n=0,\dots,N-2) \nonumber \\
\hat{q}_{N-1}^{(1)}(x) &:=&  q_{N-1}^{(1)}(x) \nonumber \\
\hat{\Phi}_n(x)  &:=& 
 {1 \over 2} \int_{-\infty}^\infty dy \,
e^{-V(y)} {\rm sgn} (x-y)  \hat{q}_n^{(1)}(x) \quad
(n=0,\dots,N-1) \nonumber \\
 f_1^{\rm odd} & = &
\left [ \begin{array}{cc}
S_1^{\rm odd}(x,y) & I_1^{\rm odd}(x,y) \\
D_1^{\rm odd}(x,y) & S_1^{\rm odd}(y,x) \end{array} \right ] \nonumber\\
S_1^{\rm odd}(x,y) & = &
\sum_{k=0}^{(N-1)/2-1}
{e^{-V(y)} \over r_k^{(1)}} \Big (
\hat{\Phi}_{2k}(x) \hat{q}_{2k+1}^{(1)}(y) -
\hat{\Phi}_{2k+1}(x)  \hat{q}_{2k}^{(1)}(y) \Big )  +
{\hat{q}^{(1)}_{N-1}(y) \over 2  r_{(N-1)/2}^{(1)} } \nonumber \\
D_1^{\rm odd}(x,y) & = & 
{\partial \over \partial x} S_1^{\rm odd}(x,y) \nonumber \\
I_1^{\rm odd}(x,y) & = & {1 \over 2} \int_{-\infty}^{\infty}
S_1^{\rm odd}(x,z) {\rm sgn} (z-y) dz -
{1 \over
2} {\rm
sgn}(x-y) + { \hat{\Phi}_{N-1}(x) \over 2 r_{(N-1)/2}^{(1)} } 
\label{2.kos}
\end{eqnarray}
With this notation, (\ref{2.7}) applies with $ f_1$ replaced by
$ f_1^{\rm odd}$ \cite{FP95}.

\subsection*{$\beta = 4$}
For $\beta = 4$ the results are independent of the parity of $N$. Let
\begin{eqnarray}
{S}_4(x,y) & = & \sum_{m=0}^{N-1} {e^{-V(x)} \over 2 r_m^{(4)}}
\Big ( q_{2m}^{(4)}(x) {d \over dy} \Big ( e^{-V(y)} 
q_{2m+1}^{(4)}(y) \Big ) -
q_{2m+1}^{(4)}(x)  
{d \over dy} \Big ( e^{-V(y)} q_{2m}^{(4)}(y) \Big ) \Big ). \nonumber
\\
{D}_4(x,y) & = &  {\partial \over \partial x} {S}_4(x,y)
\nonumber \\
{I}_4(x,y)  & := &  - \int_x^y{S}_4(x, y')dy' \label{2.10a}
\end{eqnarray}
and set
$$
{ f}_4(x,y) : = \left[
\begin{array}{ll}
\displaystyle{{S}_4(x,y)} & \displaystyle{{I}_4(x,y)} \\
\displaystyle{{D}_4(x,y)} & \displaystyle{{S}_4(y,x)}
\end{array} \right].
$$
Then with $w_4(x) = e^{-2 V(x)}$ in (\ref{2.1}) we have \cite{NW91},
\begin{equation}
\rho_{(n)}(x_1,\dots,x_n) = {\rm Tdet}[ f_4(x_j,x_k)]_{j,k=1,\dots,n}.
\end{equation}

\subsection{Construction of the skew orthogonal polynomials}
A number of works have addressed the construction of the skew orthogonal
polynomials \cite{MM91,NW91,BN91,AV99}. 
In particular Nagao and Wadati \cite{NW91} have
succeeded in giving explicit formulas for the classical skew orthogonal
polynomials in terms of related orthogonal polynomials. More recently
Adler and van Moerbeke \cite{AV99}, generalizing some earlier work of
Br\'ezin and Neuberger \cite{BN91}, have identified a natural orthogonal
polynomial basis for the   skew orthogonal polynomials
$\{q_n(x) \}$ corresponding to the skew symmetric inner products
(\ref{2.3}) and (\ref{2.4}) and shown how to use this basis to
specify the $q_n(x)$. Here we will revise this formalism, which in the
next section will be implemented in the classical cases and the results
compared with those known from the work of Nagao and Wadati \cite{NW91}.

Following \cite{AV99}, with $V(x)$ as in (\ref{2.6'}), let
\begin{equation}\label{2.fg}
2 V'(x) = {g(x) \over f(x)}
\end{equation}
be a rational function, and introduce the operator
\begin{equation}\label{2.n}
{\mbf n} := f {d \over d x} + \Big ( {f' - g \over 2} \Big ).
\end{equation}
Furthermore, introduce the symmetric inner product
\begin{equation}\label{2.innersym}
(\phi, \psi)_2 := \int_{-\infty}^\infty e^{-2 V(x)} \phi(x) \psi (x) \,
dx.
\end{equation}
Then it is shown in \cite{AV99} that
\begin{eqnarray}\label{2.n1}
(\phi, {\mbf n} \psi)_2 & = & - ({\mbf n} \phi, \psi)_2, \nonumber \\
(\phi, {\mbf n}^{-1} \psi)_2 & = & -
\ml \phi, \psi \mg_1 \Big |_{V(x) \mapsto V(x) + \log f(x)}, \nonumber \\ 
(\phi, {\mbf n} \psi)_2 & =  &  
\ml \phi, \psi \mg_4 \Big |_{V(x) \mapsto V(x) - \log f(x)}
\end{eqnarray}
(the first of these results was noted earlier in \cite{AV95}).

The practicality of these formulas lies in the fact that with
$\{p_j(x) \}_{j=0,1,\dots}$ denoting the monic orthogonal polynomials
corresponding to the symmetric inner product (\ref{2.innersym}), the
matrix $[(p_j, {\mbf n} p_k)_2]_{j,k=0,\dots, N-1}$ (which by the first
formula in (\ref{2.n1}) is skew symmetric) has non-zero terms only in
a finite number of diagonals about the main diagonal for $f$ and $g$
polynomials in (\ref{2.fg}). In particular, for the classical polynomials
this number is precisely two, giving
\begin{equation}\label{2.N}
{\cal N} := [(p_j, {\mbf n} p_k)_2]_{j,k=1,\dots, N-1} =
\left [ \begin{array}{ccccccc}
0 & c_0 & 0 & 0 &\cdots & & \\
-c_0 & 0 & c_1 & 0 &\cdots & & \\
0 & -c_1 & 0 & c_2 &\cdots &  \\
0 & 0 & -c_2 & 0 &\cdots & & \\
\vdots &\vdots &\vdots &\vdots & \ddots & & \\
& & & & & 0 & c_{N-1} \\
 & & & & & -c_{N-1} & 0 \end{array} \right ]. 
\end{equation}
Equivalently, for $\{p_j(x)\}_{j=0,1,\dots}$ a set of classical orthogonal
polynomials,
\begin{equation}\label{2.N1}
{\mbf n} p_k(x) = - {c_k \over (p_{k+1}, p_{k+1})_2} p_{k+1}(x) +
{c_{k-1} \over (p_{k-1}, p_{k-1})_2} p_{k-1}(x).
\end{equation}

The simple structure of (\ref{2.N}) allows the classical
skew orthogonal polynomials
to be expressed in terms of their orthogonal polynomial counterparts.
Let $\{\tilde{q}_j^{(1)}(x) \}$ and  $\{\tilde{q}_j^{(4)}(x) \}$
denote the skew orthogonal polynomials corresponding to the inner products
(\ref{2.3}) and (\ref{2.4}) with $V(x) \mapsto V(x) + \log f(x)$
and $V(x) \mapsto V(x) - \log f(x)$ respectively. Similarly let
$\{\tilde{r}_j^{(1)} \}$ and $\{\tilde{r}_j^{(4)}\}$ denote the
corresponding normalizations as in (\ref{2.2}). 
We can write the matrix $[(p_j, {\mbf n}^{-1} p_k)_2 ]_{j,k=0,\dots,N-1}$
in terms of ${\cal N}$. Thus with
\begin{equation}\label{2.D}
D := {\rm diag} \Big ( (p_j,p_j)_2 \Big )_{j=0,\dots,N-1}
\end{equation}
we have
\begin{equation}\label{2.18'}
\Big ( [(p_j, {\mbf n}^{-1} p_k)_2 ]_{j,k=0,\dots,N-1}
D^{-1} {\cal N} \Big )_{jk} = - \sum_{l=0}^{N-1}
{({\mbf n}^{-1} p_j, p_l)_2 \over (p_l, p_l)_2} (p_l, {\mbf n} p_k)_2 
\end{equation}
where use has been make of the first equation in (\ref{2.n1}). 
Now according to (\ref{2.N1})
$$
 \sum_{l=0}^{N-1}  {( p_l, \mbf n p_k)_2 \over (p_l, p_l)_2}
p_l(x) = {\mbf n} p_k(x) \quad {\rm for} \quad k \ne N-1.
$$
Substituting this back  in (\ref{2.18'}) shows 
\begin{equation}\label{2.D2}
[(p_j, {\mbf n}^{-1} p_k)_2 ]_{j,k=0,\dots,N-1}
D^{-1} {\cal N} = D'
\end{equation}
where $D' = D + C_k$ with $C_k$ having all entries zero except for the final
column (the explicit value of these entries will not be required).

Next introduce transition matrices $Q^{(1)}$ and $Q^{(4)}$ such that
\begin{equation}\label{2.D3}
[\tilde{q}_j^{(s)}(x)]_{j=0,\dots,N-1} =
{Q}^{(s)} [p_j(x)]_{j=0,\dots,N-1}, \qquad s = 1 \: \: {\rm or} \: \: 4.
\end{equation}
Both these transition matrices must be lower triangular matrices with
1's along the diagonal. Then we have
\begin{eqnarray}\label{2.D4}
\tilde{R}^{(1)} & := & \Big [ \ml \tilde{q}_j^{(1)}, 
\tilde{q}_k^{(1)} \mg_1  \Big |_{V(x) \mapsto V(x) + \log f(x)}
\Big ]_{j,k=0,\dots,
N-1} \nonumber \\  & = & 
Q^{(1)} [\ml p_j, p_k \mg_1 \Big |_{V(x) \mapsto V(x) + \log f(x)}
]_{j,k=0,\dots,N-1} Q^{(1)T} 
\nonumber \\
& = & - Q^{(1)} [ (p_j, {\mbf n}^{-1} p_k)_2 ]_{j,k=0,\dots,N-1}  Q^{(1)T}
\: = \: -  Q^{(1)} D' {\cal N}^{-1} D  Q^{(1)T}
\end{eqnarray}
where in obtaining the first equality of the final line the second
formula of (\ref{2.n1}) has been used, while in the final equality
(\ref{2.D2}) has been used. Taking inverses of both sides of
(\ref{2.D4}) shows that it is equivalent to the statement that
\begin{equation}\label{2.D5}
 Q^{(1)T} \tilde{R}^{(1) \, -1} Q^{(1)} =  - D^{-1} {\cal N} {D'}^{-1}.
\end{equation}
A similar calculation shows
\begin{equation}\label{2.D6}
 \tilde{R}^{(4)} :=  \Big [ \ml \tilde{q}_j^{(4)}, 
\tilde{q}_k^{(4)} \mg_4  \Big |_{V(x) \mapsto V(x) - \log f(x)} 
\Big ]_{j,k=0,\dots,
N-1} =   Q^{(4)} {\cal N} Q^{(4)T}
\end{equation}
or equivalently
\begin{equation}\label{2.D7}
Q^{(4) \, -1} \tilde{R}^{(4)} (  Q^{(4)T} )^{-1} =  {\cal N}.
\end{equation}

\section{The classical skew orthogonal polynomials}
\setcounter{equation}{0}

\subsection{Summary of known results}

\subsection*{Hermite case}
Here $e^{-V(x)} = e^{-x^2 / 2}$. The monic orthogonal polynomials with
respect to the inner product (\ref{2.innersym}), and corresponding
normalization are then 
\begin{eqnarray}\label{3.1}
p_k(x) & = & 2^{-k} H_k(x) \nonumber \\
(p_k,p_k)_2 & = & \pi^{1/2} 2^{-k} k!.
\end{eqnarray}
The skew orthogonal polynomials and corresponding normalizations
can be expressed in terms of these
orthogonal polynomials according to \cite{MM91}
\begin{eqnarray}
q_{2m}^{(1)}(x) & = & p_{2m}(x) \nonumber \\
q_{2m+1}^{(1)}(x) & = & p_{2m+1}(x) - m p_{2m-1}(x) \nonumber \\
r_n^{(1)} & = & 2^{-2n} \sqrt{\pi} \Gamma (2n+1) \label{3.1'}
\end{eqnarray}
and
\begin{eqnarray}
q_{2m+1}^{(4)}(x) & = & p_{2m+1}(x) \nonumber \\
q_{2m}^{(4)}(x) & = & m! \sum_{n=0}^m
{1 \over n!} p_{2n}(x) \nonumber \\
r_n^{(4)} & = & 2^{-2n - 1} \sqrt{\pi} \Gamma (2n + 2).
\end{eqnarray}

\subsection*{Laguerre case}
Here we have
\begin{eqnarray}\label{3.33'}
e^{-V(x)} & = & x^{a/2} e^{-x/2} \nonumber \\
p_k(x) & = & (-1)^k k! L_k^a(x) \nonumber \\
(p_k,p_k)_2 & = & \Gamma(k+1) \Gamma(a+k+1)
\end{eqnarray}
where $L_k^a$ denotes the Laguerre polynomial. With
\begin{equation}\label{3.33}
e^{-V(x)} \mapsto e^{-V(x) - \log f(x)} = x^{(a-1)/2} e^{-x/2}
\end{equation}
in (\ref{2.3}) the corresponding monic skew orthogonal polynomials
and normalization are \cite{NW91}
\begin{eqnarray}
\tilde{q}_{2m}^{(1)}(x) & = & p_{2m}(x) \nonumber \\
\tilde{q}_{2m+1}^{(1)}(x) & = & - (2m+1)! L_{2m+1}^{a-1}(x) -
(2m)! (a+2m) {d \over dx} L_{2m}^a(x) \nonumber \\
& = & p_{2m+1}(x) + (2m+1) p_{2m}(x) - 2m(a+2m) p_{2m-1}(x)
\label{3.t2} \nonumber \\
\tilde{r}_n^{(1)} & = & 2 \Gamma(n+1) \Gamma(a+2n + 2)
\label{3.t3} 
\end{eqnarray}
where in obtaining the second equality in (\ref{3.t2}) use has been made of
the formulas
\begin{equation}\label{3.6'}
L_n^{a-1}(x) = L_n^a(x) - L_{n-1}^a(x) \quad
{d \over dx} L_n^a(x) = - L_{n-1}^{a+1}(x).
\end{equation}
Recalling the fact that the skew orthogonal polynomials are non-unique up
to the transformation (\ref{2.re}) we see that we can equally as well
write in place of the second formula in (\ref{3.t2})
\begin{equation}\label{3.t2'}
\tilde{q}_{2m+1}^{(1)}(x) = p_{2m+1}(x)  - 2m(a+2m) p_{2m-1}(x).
\end{equation}

At $\beta = 4$ put
\begin{equation}\label{3.3a}
e^{-2V(x)} \mapsto e^{- 2( V(x) - \log f(x))} = x^{a+1} e^{-x}
\end{equation}
in (\ref{2.4}). Then we have \cite{NW91}
\begin{eqnarray}
\tilde{q}_{2m+1}^{(4)}(x) & = & p_{2m+1}(x) \nonumber \\
\tilde{q}_{2m}^{(4)}(x) & = & 2^{2m} m! (a/2 + m)! \sum_{j=0}^m
{1 \over 2^{2j} j! (a/2 + j)!} p_{2j}(x) \nonumber \\
\tilde{r}_n^{(4)} & = & {1 \over 2} \Gamma (2n + 2) \Gamma (a + 2n + 2).
\label{3.aa}
\end{eqnarray}

\subsection*{Jacobi case}
In the Jacobi case
\begin{equation}
e^{-V(x)} = (1 - x)^{a/2} (1 + x)^{b/2}.
\end{equation}
The monic orthogonal polynomials 
and normalizations with respect to (\ref{2.innersym}) are then
\begin{eqnarray}\label{3.jac}
p_k(x) & = & 2^k k! {\Gamma (a + b + k +1) \over
\Gamma (a + b + 2k + 1) } P_k^{(a,b)}(x) \nonumber \\
(p_k,p_k)_2 & = & 2^{(a+b+1+2k)}
{\Gamma (k+1) \Gamma (a + b + k + 1) \Gamma (a + 1 + k) \Gamma (b + 1 + k)
\over \Gamma (a + b + 2k + 2) \Gamma (a + b + 2k + 1) }
\end{eqnarray}
With
\begin{equation}
e^{-V(x)} \mapsto e^{-V(x) - \log f(x)} = (1 - x)^{(a-1)/2} (1 + x)^{
(b-1)/2}
\end{equation}
in (\ref{2.3}) the skew orthogonal polynomials and corresponding 
normalization at $\beta = 1$ are \cite{NW91}
\begin{eqnarray}
\tilde{q}_{2m}^{(1)}(x) & = & p_{2m}(x) \nonumber  \\
\tilde{q}_{2m+1}^{(1)}(x) & = &  2^{2m+1}(2m)! 
{\Gamma(a+b+2m-1) \over \Gamma(a+b+4m+1)}
\bigg\{
(2m+1)(a+b+2m-1) P_{2m+1}^{(a-1,b-1)}(x) \nonumber \\ && - 
{(a+b-2)(a+2m)(b+2m) \over
(a+b+4m)(a+b+4m+2)} {d \over dx} P_{2m}^{(a-1,b-1)}(x) \bigg\} \nonumber \\
& = & p_{2m+1}(x) + 2(2m+1) {a-b \over 4m + a + b + 2} p_{2m}(x) \nonumber
\\ && - 8m {(2m + a)(2m+b)(a+b+2m) \over
(a+b+4m-1)(a+b+4m)(a+b+4m+1)(a+b+4m+2)} p_{2m-1}(x) \nonumber \\
\tilde{r}_n^{(1)} & = & 2^{a+b+4n+2} (2n)!
{\Gamma(a+2n+1) \Gamma(b+2n+1) \Gamma(a+b+2n+1) \over
\Gamma(a+b+4n+1) \Gamma(a+b+4n+3)} \label{3.j3}
\end{eqnarray}
where in obtaining the second equality in 
the formula for $\tilde{q}_{2m+1}(x)$
use has been made of
the formulas
$$
{d \over dx} P_n^{(\alpha, \beta)}(x)  =  {1 \over 2} P_{n-1}^{\alpha+1,
\beta+1)} (x)
$$
\begin{eqnarray*} \lefteqn{
(2n+ \alpha + \beta + 1)  P_n^{(\alpha, \beta)}(x)   =  } \nonumber \\ && 
{(n+ \alpha + \beta + 1) (n + \alpha + \beta + 2) \over
(2n + \alpha + \beta + 2)} P_n^{(\alpha + 1, \beta + 1)}(x) + \nonumber \\
&&  \bigg ( - {(n + \alpha + \beta + 1) (n + \beta + 1)
\over (2n + \alpha + \beta + 2)} +
{(n+\alpha + \beta + 1) (n + \alpha) \over (2n + \alpha + \beta)}
\bigg ) P_{n-1}^{(\alpha + 1,\beta + 1)}(x) \nonumber \\&& 
- {(n+\alpha)(n+\beta) \over (2n+\alpha+\beta)}
P_{n-2}^{(\alpha+1,\beta+1)}(x).
\end{eqnarray*}
The non-uniqueness of the skew orthogonal polynomials up to the transformation
(\ref{2.re}) implies we can equally as well write $\tilde{q}_{2m+1}(x)$ 
in (\ref{3.j3}) as
\begin{eqnarray}
\lefteqn{\tilde{q}_{2m+1}(x) 
 =  p_{2m+1}(x) } \nonumber \\ &&
 - 8m {(2m + a)(2m+b)(a+b+2m) \over
(a+b+4m-1)(a+b+4m)(a+b+4m+1)(a+b+4m+2)} p_{2m-1}(x).
\end{eqnarray}

At $\beta = 4$, make the replacement
\begin{equation}
e^{-2V(x)} \mapsto e^{-2(V(x) - \log f(x) )} = (1-x)^{a+1} (1+x)^{b+1}
\end{equation}
in (\ref{2.4}). The corresponding
skew orthogonal polynomials and normalizations are then \cite{NW91}
\begin{eqnarray}
\tilde{q}_{2m+1}^{(4)}(x) & = & p_{2m+1}(x) \nonumber \\
\tilde{q}_{2m}^{(4)}(x) & = & {2^{6m+a+b} \over \sqrt{\pi}}
{m! \Gamma (a/2 + b/2 + m +1) \Gamma (a/2 + m + 1)
\Gamma (b/2 + m +1) \over \Gamma (a + b + 4m + 2) } \nonumber
\\ && \times \sum_{j=0}^m {1 \over j! 2^{4j}}
{\Gamma ((a/2 + b/2 + j + 1)
\over \Gamma (a/2 + j + 1) \Gamma (b/2 + j + 1)}
{ \Gamma(a + b + 4j + 2) \over \Gamma (a+b+2j+1)} p_{2j}(x)
\nonumber \\ &= &  2^{6m} m!
{\Gamma (a/2 + b/2 + m + 1) \Gamma (a/2 + m + 1) \Gamma (b/2 + m + 1)
\over \Gamma (a + b + 4m + 2) } \nonumber 
\\ &&\times \sum_{j=0}^m {1 \over j! 2^{6j}}
{\Gamma (a + b + 4j + 2) \over
\Gamma (a/2 + b/2 + j + 1)   \Gamma (a/2 + j + 1) \Gamma (b/2 + j + 1)}
p_{2j}(x) \label{3.2j2} \nonumber \\
\tilde{r}_n^{(4)} & = &
{2^{a+b+4n+2}\Gamma(2n+2)\Gamma(a+2n+2)\Gamma(b+2n+2)\Gamma(a+b+2n+2) \over
\Gamma(a+b+4n+2)\Gamma(a+b+4n+4)} \label{3.2j3}
\end{eqnarray}
where in obtaining the second equality in the formula for
$\tilde{q}_{2m}^{(4)}(x)$  use has been made of
the duplication formula
$$
\Gamma (z) \Gamma (z + 1/2) = 2^{1/2 - 2z} (2 \pi )^{1/2} \Gamma (2z).
$$

\subsection{Derivation from general formalism}
\subsection*{$\beta = 1$}
The expression for skew orthogonal polynomials at $\beta = 1$ is determined
by solving the equation (\ref{2.D5}) for $Q^{(1)} = [Q_{jk}]_{j,k=0,\dots,N-1}$.
In fact this can be done by premultiplying both sides by
$( Q^{(1)T})^{-1}$, and equating the lower triangular entries on
both sides (not including the diagonal). Using (\ref{2.2}) we see that on
the l.h.s.
\begin{equation}\label{3.lhs1}
\Big ( \tilde{R}^{-1} Q^{(1)} \Big )_- = 
\left ( \begin{array}{ccccccc}
* & & & & & &\\
1/\tilde{r}_0^{(1)} & * & & & & & \\
-Q_{30} / \tilde{r}_1^{(1)} & - Q_{31} / \tilde{r}_1^{(1)} &
* & & &  & \\
Q_{20}/ \tilde{r}_1^{(1)} & Q_{21} / \tilde{r}_1^{(1)} &
1/\tilde{r}_1^{(1)} & * & & & \\
- Q_{50} / \tilde{r}_2^{(1)} & - Q_{51} / \tilde{r}_2^{(1)} &
- Q_{52} / \tilde{r}_2^{(1)} & - Q_{53} / \tilde{r}_2^{(1)} & * &  & \\
Q_{40} / \tilde{r}_2^{(1)} & Q_{41} / \tilde{r}_2^{(1)} &
 Q_{42} / \tilde{r}_2^{(1)} & Q_{43} / \tilde{r}_2^{(1)} &
1/\tilde{r}_2^{(1)} & * & \\
\vdots&\vdots &\vdots & \vdots &\vdots & 
& \ddots \end{array} \right )_{2N \times 2N}
\end{equation} 
where the subscript in $(\,)_-$ denotes the strictly lower triangular entries.
On the r.h.s.~we note from the fact that $( Q^{(1)T})^{-1}$ is
an upper triangular matrix with 1's along the diagonal and the explicit
form (\ref{2.N}) of ${\cal N}$ that
\begin{eqnarray}\label{3.rhs1}
- \Big ( ( Q^{(1)T})^{-1} D^{-1} {\cal N} {D'}^{-1} \Big )_- & = &
- \Big ( D^{-1} {\cal N} {D'}^{-1} \Big )_- \nonumber \\
& = &
\left ( \begin{array}{ccccc}
* & & & & \\
 \gamma_0 & * & & & \\
0 &  \gamma_1 & * & & \\
0 & 0 &  \gamma_2 & * & \\
\vdots & \vdots & \vdots &  & \ddots \end{array} \right )
\end{eqnarray}
where
\begin{equation}\label{3.gam}
\gamma_j := c_j / (p_{j+1}, p_{j+1})_2 (p_j,p_j)_2.
\end{equation}
Equating (\ref{3.lhs1}) and (\ref{3.rhs1}) gives
\begin{eqnarray}
\gamma_{2p} & = &  1/ \tilde{r}_p^{(1)} \quad
(p=0,\dots,N-1) \nonumber \\
{Q}_{2p,l} & = & 0 \quad (l=0,\dots,2p-1)  \nonumber \\
{Q}_{2p+1,l} & = & 0 \quad (l=0,\dots,2p-2)  \nonumber \\
{Q}_{2p+1,2p-1} & = & - \gamma_{2p-1} \tilde{r}_p^{(1)} =
 - \gamma_{2p-1} / \gamma_{2p}
\end{eqnarray}
while ${Q}_{2p+1,2p}$ is left unspecified. Hence
\begin{eqnarray}
\tilde{q}_{2j}^{(1)}(x) & = & p_{2j}^{(1)}(x)  \nonumber \\
\tilde{q}_{2j+1}^{(1)}(x) & = &  p_{2j+1}^{(1)}(x) +
{Q}_{2p+1,2p} p_{2j}^{(1)}(x) -
{\gamma_{2j-1} \over \gamma_{2j}} p_{2j-1}^{(1)}(x) \nonumber \\
\tilde{r}_p^{(1)} & = & - 1/ \gamma_{2p}. \label{3.p13}
\end{eqnarray}
Note that the fact that ${Q}_{2p+1,2p}$ is arbitrary in the formula
for $\tilde{q}_{2j+1}^{(1)}(x)$ is
consistent with the skew orthogonal polynomials being non-unique up
to the transformation (\ref{2.re}). The simplest choice is 
${Q}_{2p+1,2p} = 0$.

Let us verify that (\ref{3.p13}) with ${Q}_{2p+1,2p} = 0$
reclaims the results
(\ref{3.1}), (\ref{3.t3}) and (\ref{3.j3}). First $\gamma_j$ must be specified
(recall (\ref{3.gam})). From (\ref{2.N1}) and (\ref{3.gam}) we have that
\begin{equation}\label{3.nn}
{\mbf n} p_k(x) = - \gamma_k (p_k,p_k)_2 \, p_{k+1}(x) + 
{\rm lower \, degree \, term},
\end{equation}
which together with the explicit form of ${\mbf n}$ 
for each of the classical weight functions and the fact that
each $p_k(x)$ is monic implies
\begin{equation}\label{3.hlj}
\gamma_k (p_k, p_k)_2 = \left \{\begin{array}{ll}
1, & {\rm Hermite} \\[.2cm]
{1 \over 2}, & {\rm Laguerre} \\[.2cm]
 {1 \over 2} (2k + 2 + a + b), & {\rm Jacobi} \end{array} \right.
\end{equation}
Furthermore, the explicit form of $(p_k,p_k)_2$ is given by
(\ref{3.1}), (\ref{3.33'}) and (\ref{3.jac}) in the Hermite, Laguerre and
Jacobi cases respectively. A straightforward
calculation then enables the formulas of the previous
section for $\beta = 1$ to be reclaimed from the general formulas
(\ref{3.p13}).

\subsection*{$\beta = 4$}
At $\beta = 4$ we must solve the equation (\ref{2.D7}). This we will do
for the matrix $(Q^{(4)})^{-1} := L = [\tilde{\beta}_{jk}^{(4)}]_{j,k=0,\dots,
N-1}$, where $L$ is lower triangular with
1's along the diagonal. In this case it is sufficient to multiply
both sides of (\ref{2.D7}) by $Q^{(4)T}$ and equate the strictly lower
triangular parts of both sides. On the l.h.s.~we have
\begin{equation}\label{3.lhs3}
\Big ( L \tilde{R}^{(4)} \Big )_- =
\left ( \begin{array}{cccccc}
* & & & & & \\
-\tilde{r}_0^{(4)} & * & & & & \\
-\tilde{r}_0^{(4)} \tilde{\beta}_{21}^{(4)}
& \tilde{r}_0^{(4)} \tilde{\beta}_{20}^{(4)}
& * & & & \\
-\tilde{r}_1^{(4)} \tilde{\beta}_{31}^{(4)} & 
\tilde{r}_1^{(4)} \tilde{\beta}_{30}^{(4)} &
-  \tilde{r}_1^{(4)} & * & & \\
-\tilde{r}_1^{(4)}\tilde{\beta}_{41}^{(4)} & 
\tilde{r}_1^{(4)} \tilde{\beta}_{40}^{(4)} &
-\tilde{r}_1^{(4)} \tilde{\beta}_{43}^{(4)} & 
\tilde{r}_1^{(4)} \tilde{\beta}_{42}^{(4)} & * & \\
\vdots & \vdots & \vdots & \vdots & & \ddots
\end{array} \right )_{2N \times 2N}
\end{equation}
while on the r.h.s.
\begin{equation}\label{3.rhs3}
 \Big ( N Q^{(4)T} \Big )_- =   ( N )_-.
\end{equation}
Equating (\ref{3.lhs3}) and (\ref{3.rhs3}) gives
\begin{eqnarray}
c_{2p} & = & \tilde{r}_p^{(4)} \qquad  \: \: p=0,\dots, N - 1 \nonumber \\
\tilde{\beta}_{2p+1,j}^{(4)} & = & 0 \qquad \qquad j=0,\dots, 2p-1 \nonumber \\
\tilde{\beta}_{2p, j}^{(4)} 
& = & 0 \qquad \qquad j=0,\dots, 2p-3, \: j=2p-1 \nonumber \\
\tilde{\beta}_{2p, 2p-2}^{(4)}
 & = & - c_{2p-1} / \tilde{r}_{p-1}^{(4)} = - c_{2p-1} / c_{2p-2}
\end{eqnarray}
while $\tilde{\beta}_{2p+1, 2p}^{(4)}$ is undetermined. Thus we have
\begin{eqnarray}
p_{2j+1}(x) & = & \tilde{q}_{2j+1}^{(4)}(x) + \tilde{\beta}_{2j+1,2j}^{(4)}
 \tilde{q}_{2j}^{(4)}(x) \nonumber \\
p_{2j}(x) & = & \tilde{q}_{2j}^{(4)}(x) - {c_{2j-1} \over c_{2j-2}}
\tilde{q}_{2j-2}^{(4)}(x). \nonumber \\
\tilde{r}_p^{(4)} & = &  c_{2p}. \label{3.gc}
\end{eqnarray}

The non-uniqueness of the skew orthogonal polynomials up to the
transformation (\ref{2.re}) implies we can choose 
$\tilde{\beta}_{2j+1,2j}^{(4)} = 0$.
Doing this, and solving for $\tilde{q}_{2j}^{(4)}(x)$ in the second equation
of (\ref{3.gc}) gives
\begin{eqnarray}
 \tilde{q}_{2j+1}^{(4)}(x) & = & p_{2j+1}(x) \nonumber \\
 \tilde{q}_{2j}^{(4)}(x)  & = & 
\Big ( \prod_{p=0}^{j-1} {c_{2p+1} \over c_{2p}} \Big )
\sum_{l=0}^j \Big ( \prod_{p=0}^{l-1}
{c_{2p+1} \over c_{2p}} \Big )^{-1} p_{2l}(x) \nonumber \\
\tilde{r}_p^{(4)} & = & c_{2p}. \label{3.gc1}
\end{eqnarray}
We recall that the quantities $c_j$ are, according to (\ref{3.gam}) and
(\ref{3.hlj}), simply related to the norm $(p_{j+1}, p_{j+1})_2$
which in turn is given by (\ref{3.1}), (\ref{3.33'}) and (\ref{3.jac})
in the Hermite, Laguerre and Jacobi cases respectively. From this it is
straightforward to verify that the general formula reproduces the results
of the previous section for $\beta = 4$.

\section{Summation formulas}
\setcounter{equation}{0}
The results of Section \ref{s2.1} show that the $n$-point distributions are
given in terms of the quantities $S_1(x,y)$ 
and $S_4(x,y)$ 
 (recall (\ref{2.8a}) and (\ref{2.10a}))
at $\beta = 1$ and $\beta = 4$ respectively.
It is of interest to compare these formulas with the expression for the
$n$-point distribution (\ref{2.1}) at $\beta = 2$. With $w_2(x) =
e^{-2V(x)}$ it is an easy consequence of (\ref{1.2}) that
\begin{equation}
\rho_{(n)}(x_1,\dots,x_n) = \det [ S_2(x_j,x_k) ]_{j,k=1,\dots,n}, \quad
S_2(x,y) : =  e^{-V(x)-V(y)} \sum_{l=0}^{N-1}
{p_l(x) p_l(y) \over (p_l, p_l)_2 }. 
\end{equation}
Thus again the $n$-point function is completely determined by a summation.
In fact for $\{p_j(x) \}_{j=0,1,\dots}$ a general set of monic orthogonal
polynomials this sum can be evaluated exactly according to the
Christoffel--Darboux formula
\begin{equation}\label{3.cd}
 \sum_{l=0}^{N-1} {p_l(x) p_l(y) \over (p_l, p_l)_2 } =
{1 \over (p_{N-1}, p_{N-1})_2 }
{p_N(x) p_{N-1}(y) - p_{N-1}(x) p_N(y) \over x - y}.
\end{equation}
Here we seek an analogous summation formula for $S_1(x,y)$ and $S_4(x,y)$.

There is some recent literature on this problem. Using different methods,
Widom \cite{Wi98} and Forrester et al.~\cite{FNH98} have expressed
$S_1(x,y)$ and $S_4(x,y)$ in terms of the Christoffel--Darboux
summation (\ref{3.cd}) in which $\{p_j(x)\}$ are the monic orthogonal
polynomials associated with the weight function $w_2(x) = e^{-2V(x)}$,
plus a `correction' term. In general the correction term (which is given
in a different form in the two different formalisms
\cite{Wi98} and \cite{FNH98}) does not appear to have a closed form
evaluation. Exceptions are the Hermite (H) and Laguerre (L) cases.
Then the correction term factorizes as  a function of $x$ times a
function of $y$, both of which can be computed exactly (the results
in the Hermite case were known from earlier work; see e.g.~\cite{Me91}). 
Explicitly
\begin{eqnarray}
S_1^{(G)}(x,y) & = & S_2^{(G)}(x,y) +
{e^{-y^2/2} \over 2^{N+1} \sqrt{\pi} (N-1)! } H_{N-1}(y) 
\int_{-\infty}^\infty {\rm sgn}(x-t) e^{-t^2/2} H_N(t) \, dt
\nonumber \\
S_1^{(G){\rm odd}}(x,y) & = & S_1^{(G)}(x,y) +
{H_{N-1}(x) \over \int_{-\infty}^\infty e^{-t^2/2} H_{N-1}(t) \, dt}
\nonumber \\
2 S_4^{(G)}(x,y) & = & S_2^{(G)}(x,y)  \Big |_{N \mapsto 2N}
+  e^{-y^2/2} H_{2N}(y)
{2^{-2N} \over \sqrt{\pi} \Gamma (2N) }
\int_{-\infty}^x e^{-t^2 / 2} H_{2N-1}(t) \, dt
  \nonumber \\
S_1^{(L)}(x,y) & = & S_2^{(L)}(x,y) + {N! \over 4 \Gamma (N + a)} 
y^{a/2} e^{-y/2} \nonumber \\
&& \times \Big ( {d \over d y} L_N^a(y) \Big )
\int_0^\infty {\rm sgn} (x-y) y^{a/2 - 1} e^{-y/2}
\Big ( L_N^a(y) - L_{N-1}^a(y) \Big ) \, dy \nonumber \\
2 S_4^{(L)}(x,y)  & = &  S_2^{(L)}(x,y) \Big |_{N \mapsto 2N} \nonumber \\
&& + {(2N)! y^{a/2} e^{-y/2} \over 2 \Gamma (2N + a)}
{L_{2N}^a(y) - L_{2N-1}^a(y) \over y}
\int_0^x t^{a/2} e^{-t/2} {d \over dt} L_{2N}^a(t) \, dt. 
 \label{4.tim}
\end{eqnarray} 

One feature of these formulas is that the quantities $S_1(x,y)$ and
$S_4(x,y)$, in which the skew orthogonal polynomials are
with respect to the skew inner products (\ref{2.3}) and (\ref{2.4})
(and thus involve $e^{-V(x)}$ and $e^{-2V(x)}$ respectively) are expressed
in terms of the weight function $e^{-2V(x)}$. However, we have seen in
the previous section that it is the skew orthogonal polynomials
at $\beta = 1$ with the replacement
\begin{equation}\label{4.R1}
V(x) \mapsto V(x) + \log f(x)
\end{equation}
in (\ref{2.3}), and the  skew orthogonal polynomials at $\beta = 4$ with
the replacement
\begin{equation}\label{4.R2}
V(x) \mapsto V(x) - \log f(x)
\end{equation}
in (\ref{2.4}) which are naturally expressed in terms of the orthogonal 
polynomials for the weight function $e^{-2V(x)}$. This suggests that
we consider the quantities $\tilde{S}_1(x,y)$ and $\tilde{S}_4(x,y)$,
defined as in (\ref{2.8a}) and (\ref{2.10a}) but with the 
corresponding skew
orthogonal polynomials  defined with the replacements
(\ref{4.R1}) and (\ref{4.R2}) respectively, and seek to sum them in terms
of the orthogonal polynomials for the weight function $e^{-2 V(x)}$.
In the Hermite case nothing changes because, according to the definition
(\ref{2.fg}), $f(x) = 1$ and so $\log f(x) = 0$ in (\ref{4.R1}) and
(\ref{4.R2}). However, in the Laguerre case this is already a different
viewpoint as then $V(x)$ is modified by $a \mapsto a \pm 1$. Indeed we
find that the simple structures (\ref{3.p13}) and (\ref{3.gc}) giving the
skew orthogonal polynomials for the modified $V(x)$ in terms of the
orthogonal polynomials for the weight function $e^{-2V(x)}$ allows
$\tilde{S}_1(x,y)$ and $\tilde{S}_4(x,y)$ to be summed in the classical
cases by the same general formula in each case.

\subsection*{$\beta = 1$, $N$ even}
Let us write
\begin{equation}\label{4.p1}
p_l(x) = \sum_{j=0}^l \tilde{\beta}_{lj}^{(1)} \tilde{q}_j^{(1)}(x), \quad
\tilde{\beta}_{ll}^{(1)} = 1
\end{equation}
where $\{  \tilde{q}_j^{(1)}(x) \}_{j=0,1,\dots}$ is a set of monic skew
 orthogonal polynomials with respect to the inner product (\ref{2.3})
(modified so that $V$ is replaced by $\tilde{V}$), and
$\{ p_l(x) \}_{l=0,1,\dots}$ is the set of monic polynomials with
respect to the inner product (\ref{2.innersym}). The quantity
$S_1(x,y)$ can be expressed in terms of the polynomials $p_l(x)$.

\begin{prop}
Let $N$ be even. With $\tilde{S}_1(x,y)$ defined by (\ref{2.8a}),
modified so that $V \mapsto \tilde{V}$ and $q_k^{(1)}(x) \mapsto
\tilde{q}_k^{(1)}(x)$, and $\{p_j(x)\}_{j=0,1,\dots}$ the set of
monic orthogonal polynomials associated with the weight function
$e^{-2V(x)}$ (assumed complete) we have
\begin{equation}\label{4.it}
\tilde{S}_1(x,y) = e^{-2 V(x) + \tilde{V}(x) - \tilde{V}(y) }
\bigg ( \sum_{n=0}^{N-1} {p_n(x) p_n(y) \over (p_n,p_n)_2 }
+ \sum_{n=N}^\infty \sum_{k=0}^{N-1} {p_n(x) \over (p_n,p_n)_2}
\tilde{\beta}_{nk}^{(1)} \tilde{q}_k^{(1)}(y) \bigg ).
\end{equation}
\end{prop}

\noindent{\bf Proof} \quad Our derivation is motivated by the derivation
of a similar formula in \cite{NF98}. Now, from the definitions
(\ref{2.Phi}) and (\ref{2.3}) (the latter with $V \mapsto \tilde{V}$) it
is possible to write
$$
\tilde{\Phi}_k(x) = 
e^{\tilde{V}(x)} \ml \tilde{q}_k^{(1)}(y) | \delta (x-y) \mg_1.
$$
But from the completeness of $\{p_j(x) \}$ we can make the expansion
\begin{equation}\label{4.ex}
\delta (x - y) = e^{-2 \tilde{V} (x) } \sum_{n=0}^\infty
{p_n(x) p_n(y) \over (p_n, p_n)_2 }.
\end{equation}
Substituting (\ref{4.p1}) for $p_n(y)$ and using the skew orthogonality
of $\{\tilde{q}_k(y)\}$ shows
\begin{eqnarray}
\tilde{\Phi}_{2k}(x) & = & \tilde{r}_k^{(1)} e^{\tilde{V}(x)} e^{-2 V(x)}
\sum_{\nu = 2k + 1}^\infty {p_\nu(x) \over (p_\nu, p_\nu)_2}
\tilde{\beta}_{\nu \, 2k + 1}^{(1)} \label{4.8} \\
\tilde{\Phi}_{2k+1}(x) & = & - \tilde{r}_k^{(1)} e^{\tilde{V}(x)} e^{-2 V(x)}
\sum_{\nu = 2k}^\infty {p_\nu(x) \over (p_\nu, p_\nu)_2}
\tilde{\beta}_{\nu \, 2k}^{(1)} \label{4.8a}
\end{eqnarray}
Next substitute these formulas in the definition (\ref{2.8a}). The stated
formula follows after minor manipulation involving further use of
(\ref{4.p1}).

\vspace{.2cm}
The first sum in (\ref{4.it}) is evaluated according to the
Christoffel-Darboux formula (\ref{3.cd}). This applies for general monic
orthogonal polynomials $\{p_j(x)\}$. To evaluate the second sum requires
knowledge of the transition coefficients $\tilde{\beta}_{lj}^{(1)}$. In the
case of the classical polynomials these coefficients can be determined
from (\ref{3.p13}). Setting $Q_{2p+1,2p} = 0$ and comparing with the
definition (\ref{4.p1}) we see that
\begin{eqnarray*}
\tilde{\beta}_{2l,j}^{(1)} & = & 0, \quad (j=0,\dots,2l-1) \\
\tilde{\beta}_{2l+1,2j}^{(1)}  & = & 0, \quad (j =0,\dots, l ) \\
\tilde{\beta}_{2l+1,2j+1}^{(1)} & = & 
{\prod_{k=1}^l a_k \over \prod_{k=1}^j a_k},
\quad a_k := {\gamma_{2k-1} \over \gamma_{2k}}.
\end{eqnarray*}
In fact we don't require the explicit form of the $\tilde{\beta}_{lj}^{(1)}$,
but rather their factorization property
\begin{equation}\label{4.fac}
\tilde{\beta}_{nk}^{(1)} = \tilde{\beta}_{n,N-1}^{(1)}
 \tilde{\beta}_{N-1,k}^{(1)}, \quad n \ge N,
\end{equation}
which is evident from the above formulas. Substituting (\ref{4.fac}) in the
double summation in (\ref{4.it}) and recalling (\ref{4.p1}) shows
\begin{eqnarray*}
\sum_{n=N}^\infty \sum_{k=0}^{N-1} {p_n(x) \over (p_n,p_n)_2}
\tilde{\beta}_{nk}^{(1)} \tilde{q}_k(y) & = &
\Big ( \sum_{n=N}^\infty {p_n(x) \over (p_n,p_n)_2}
\tilde{\beta}_{n, N-1}^{(1)} \Big ) p_{N-1}(y)  \\
& = & \bigg ( {1 \over \tilde{r}_{N/2 - 1}^{(1)}}
e^{-V(x) + 2 \tilde{V}(x)} \tilde{\Phi}_{N-2}(x) -
{p_{N-1}(x) \over (p_{N-1}, p_{N-1})_2} \bigg ) p_{N-1}(y)
\end{eqnarray*}
where the second equality follows from (\ref{4.8}) and the fact that $N$
is assumed even. Thus
\begin{eqnarray}\label{4.sum1}
\lefteqn{
\tilde{S}_1(x,y)   } \nonumber \\
&&  = e^{-({V}(x) - \tilde{V}(x))} 
e^{(V(y) - \tilde{V}(y))}
S_2(x,y) \Big |_{N \mapsto
N-1} + \gamma_{N-2} e^{-\tilde{V}(y)} 
\tilde{\Phi}_{N-2}(x) p_{N-1}(y) \nonumber \\
&  & = e^{-({V}(x) - \tilde{V}(x))} e^{(V(y) - \tilde{V}(y))}  
S_2(x,y) \Big |_{N \mapsto
N-1}  \nonumber \\ &&  +  \gamma_{N-2} e^{-\tilde{V}(y)} p_{N-1}(y)
{1 \over 2} \int_{-\infty}^\infty {\rm sgn}(x-t) p_{N-2}(t)
e^{-\tilde{V}(t)} \, dt, \nonumber \\
\end{eqnarray}
where $\gamma_{N-2}$ is specified in terms of $(p_{N-2}, p_{N-2})_2$
by (\ref{3.hlj}). Hence the summation of $\tilde{S}_1$ takes the
same form for all the classical cases.

It is of interest to compare (\ref{4.sum1}) with the summation formulas
for $\tilde{S}_1$ in (\ref{4.tim}). In fact neither the Hermite nor
Laguerre summations in (\ref{4.tim}) are of the same form as
(\ref{4.sum1}). But in both cases the equality of the different forms
can be established directly. Consider (\ref{4.sum1}) in the Hermite case
when it reads
$$
\tilde{S}_1^{(G)}(x,y)  =  S_2^{(G)}(x,y) \Big |_{N \mapsto N -1}
+ {e^{-y^2/2} H_{N-1}(y) \over 2^N \pi^{1/2} (N - 2)! }
\int_{-\infty}^\infty dt \, {\rm sgn} (x-t) e^{-t^2/2}
H_{N-2}(t).
$$
This can be brought into agreement with the formula for
$\tilde{S}_1^{(G)}$ in (\ref{4.tim}) if we note from the identity
$$
H_{2k}(y) - {1 \over 4 (k + 1/2)} H_{2k+2}(y) =
{1 \over 2k + 1} e^{y^2/2} {d \over dy} \Big (
e^{-y^2/2} H_{2k+1}(y) \Big )
$$
that
\begin{equation}\label{4.br}
\Phi_{2k}(x) - {1 \over k + 1/2} \Phi_{2k+2}(x) = {2^{-2k} \over
2 k + 1} e^{-x^2 / 2} H_{2k+1}(x).
\end{equation}
In the Laguerre case the manipulation of (\ref{4.sum1}) into the
form given in (\ref{4.tim}) is undertaken in the Appendix.

\subsection*{$\beta = 1$, $N$ odd}
In the case of $N$ odd, we see from the definitions (\ref{2.kos}) that
\begin{eqnarray*}
\tilde{S}_1^{\rm odd}(x,y) & = & \tilde{S}_1(x,y) \Big |_{N \mapsto
N-1} + {\tilde{q}_{N-1}^{(1)}(y) \over 2 \tilde{s}_{N-1}} \\
&& + {\tilde{\Phi}_{N-1}(x) \over \tilde{s}_{N-1}}
\sum_{k=0}^{(N-1)/2-1} {e^{-\tilde{V}(y)} \over \tilde{r}_k^{(1)}}
\Big ( - \tilde{s}_{2k} \tilde{q}_{2k+1}(y) + \tilde{s}_{2k+1}
\tilde{q}_{2k}(y) \Big ) \\
&& - {\tilde{q}_{N-1}^{(1)}(y) \over \tilde{s}_{N-1}}
\sum_{k=0}^{(N-1)/2-1} {e^{-\tilde{V}(y)} \over \tilde{r}_k^{(1)}}
\Big ( - \tilde{s}_{2k} \tilde{\Phi}_{2k+1}(x) + \tilde{s}_{2k+1}
\tilde{\Phi}_{2k}(x) \Big )
\end{eqnarray*}
where
\begin{equation}\label{4.s}
\tilde{s}_k := {1 \over 2} \int_{-\infty}^\infty
e^{-\tilde{V}(x)} \tilde{q}_k^{(1)}(x) \, dx.
\end{equation}
The quantity $ \tilde{S}_1(x,y) \Big |_{N \mapsto
N-1}$ is evaluated by (\ref{4.sum1}). Furthermore, from the definitions
we see that
\begin{eqnarray}\lefteqn{
\sum_{k=0}^{(N-1)/2-1} {e^{-\tilde{V}(y)} \over \tilde{r}_k^{(1)}}
\Big ( - \tilde{s}_{2k} \tilde{q}_{2k+1}(y) + \tilde{s}_{2k+1}
\tilde{q}_{2k}(y) \Big )} \nonumber \\  &&  = - \lim_{x \to \infty}
\tilde{S}_1(x,y) \Big |_{N \mapsto N - 1}  =
- \gamma_{N-3} \tilde{s}_{N-3} e^{-\tilde{V}(y)} p_{N-2}(y) \label{4.k1}
\end{eqnarray}
where here the second equality follows from (\ref{4.sum1}), while
\begin{eqnarray}\lefteqn{
\sum_{k=0}^{(N-1)/2-1} {1 \over \tilde{r}_k^{(1)}}
\Big ( - \tilde{s}_{2k} \tilde{\Phi}_{2k+1}(x) + \tilde{s}_{2k+1}
\tilde{\Phi}_{2k}(x) \Big )} \nonumber \\  && =
{1 \over 2} \int_{-\infty}^\infty {\rm sgn}(x-y)
\Big ( - \lim_{x' \to \infty} \tilde{S}_1(x',y) \Big |_{N \mapsto N - 1}
\Big ) \, dy 
= - \gamma_{N-3} \tilde{s}_{N-3} \tilde{\phi}_{N-2}(x)
\end{eqnarray}
where here  the second equality follows from (\ref{4.k1}) and
\begin{equation}\label{4.k2}
\tilde{\phi}_j(x) := {1 \over 2} \int_{-\infty}^\infty
e^{-\tilde{V}(y)} {\rm sgn}(x-y) p_{j}(y) \, dy.
\end{equation}
Hence
\begin{eqnarray}\label{4.sum2}
\tilde{S}_1^{\rm odd}(x,y) & = & 
\tilde{S}_1(x,y) \Big |_{N \mapsto N - 1} +
{{p}_{N-1}(y) \over 2 \tilde{s}_{N-1}} \nonumber \\
&&
- \gamma_{N-3} \tilde{s}_{N-3} {e^{-\tilde{V}(y)} \over
\tilde{s}_{N-1}} \Big ( \tilde{\phi}_{N-1}(x) p_{N-2}(y) 
- p_{N-1}(y) \tilde{\phi}_{N-2}(x)
\Big ),
\end{eqnarray}
where the fact that $\tilde{q}_{N-1}^{(1)}(y) = p_{N-1}(y)$, which follows from
(\ref{3.p13}) and the fact that $N$ is odd, has been used.

Let us show how to write (\ref{4.sum2}) in the Hermite case as presented
in (\ref{4.tim}). Now, from the definite integral 
$$
\int_{-\infty}^\infty e^{-x^2 / 2} H_n(x) \, dx =
(2 \pi)^{1/2} 2^{n/2} (n-1) (n-3) \cdots 3 \, \cdot \, 1 \qquad
n \: \: {\rm even}
$$
we see that
\begin{equation}\label{4.star}
\gamma_{N-3} {\tilde{s}_{N-3} \over \tilde{s}_{N-1}} =
2 { \gamma_{N-3} \over N - 2} = \gamma_{N-2}.
\end{equation}
This fact shows that the final term in the second line of (\ref{4.sum2})
is equal to the final term in the formula (\ref{4.sum1}) for
$\tilde{S}_1(x,y)$. Also, making use of the first equality in (\ref{4.star})
and (\ref{4.br}) shows
$$
e^{-\tilde{V}(y)} p_{N-2}(y) \gamma_{N-3}
\Big ( \tilde{\phi}_{N-3}(x) - {s_{N-3} \over s_{N-1}} 
\tilde{\phi}_{N-1}(x) \Big )
= e^{-\tilde{V}(x) - \tilde{V}(y)} {p_{N-2}(x) p_{N-2}(y) \over
(p_{N-2}, p_{N-2})_2 }.
$$
We see from (\ref{4.sum1}) with $N \mapsto N - 1$ that this identity
provides the final step in identifying (\ref{4.sum2}) in the Hermite
case with the formula in (\ref{4.tim}).

\subsection*{$\beta = 4$}
The strategy for obtaining a summation formula at $\beta = 4$ is similar
to that used at $\beta = 1$. We write
\begin{equation}\label{4.b2}
p_l(x) = \sum_{j=0}^l \tilde{\beta}_{lj}^{(4)} \tilde{q}_j(x), \quad
 \tilde{\beta}_{ll}^{(4)} = 1
\end{equation}
where $\{ \tilde{q}_j^{(4)}(x) \}_{j=0,1,\dots}$ is a set of monic skew
orthogonal polynomials with respect to the inner product (\ref{2.4})
(modified so that $V$ is replaced by $\tilde{V}$) and
$\{p_l(x)\}_{l=0,1,\dots}$ is the set of monic polynomials with respect
to the inner product (\ref{2.innersym}).

\begin{prop}
With $\tilde{S}_4(x,y)$ defined by (\ref{2.10a}),
modified so that $V \mapsto \tilde{V}$ and $q_k^{(4)}(x) \mapsto
\tilde{q}_k^{(4)}(x)$, and $\{p_j(x)\}_{j=0,1,\dots}$ the set of
monic orthogonal polynomials associated with the weight function
$e^{-2V(x)}$ (assumed complete) we have
\begin{equation}\label{4.it1}
\tilde{S}_4(x,y) = {1 \over 2} e^{-2 V(y) - \tilde{V}(x) + \tilde{V}(y) }
\bigg ( \sum_{n=0}^{2N-1} {p_n(x) p_n(y) \over (p_n,p_n)_2 }
+ \sum_{n=2N}^\infty \sum_{k=0}^{2N-1} {p_n(x) \over (p_n,p_n)_2}
\tilde{\beta}_{nk}^{(4)} \tilde{q}_k(x) \bigg ).
\end{equation}
\end{prop}

\noindent{\bf Proof}
Again our derivation is motivated by the workings in \cite{NF98}. From
the fact that
$$
{d \over dx} \Big ( e^{-\tilde{V}(x)} \tilde{q}_{2m}^{(4)}(x) \Big )
= {e^{\tilde{V}(x)} \over 2} \bigg (
 e^{-2 \tilde{V}(x)} {d \over dx}  \tilde{q}_{2m}^{(4)}(x)
+ {d \over dx} \Big (  e^{-2 \tilde{V}(x)}  \tilde{q}_{2m}^{(4)}(x)
\Big ) \bigg )
$$
we can check from the definition (\ref{2.4}) that it's possible to write
$$
{d \over dx} \Big ( e^{-\tilde{V}(x)} \tilde{q}_{2m}^{(4)}(x) \Big )
= e^{\tilde{V}(x)} \ml \delta (x-y) |  \tilde{q}_{2m}^{(4)}(y) \mg_4.
$$
Substituting (\ref{4.ex}) for $\delta (x-y)$, and then substituting
(\ref{4.b2}) for $p_j(y)$ and making use of the skew orthogonality
of $\{q_j^{(4)}(x) \}$ shows
\begin{equation}\label{4.q4a}
{d \over dx} \Big ( e^{-\tilde{V}(x)} \tilde{q}_{2m}^{(4)}(x) \Big )
= - \tilde{r}_m^{(4)} e^{\tilde{V}(x) - 2 V(x)}
\sum_{\nu = 2m + 1}^\infty {p_\nu(x) \over (p_\nu,p_\nu)_2}
\tilde{\beta}_{\nu, 2m+1}^{(4)}.
\end{equation}
Proceeding similarly we can also show
\begin{equation}\label{4.q4b}
{d \over dx} \Big ( e^{-\tilde{V}(x)} \tilde{q}_{2m+1}^{(4)}(x) \Big )
=  \tilde{r}_m^{(4)} e^{\tilde{V}(x) - 2 V(x)}
\sum_{\nu = 2m}^\infty {p_\nu(x) \over (p_\nu,p_\nu)_2}
\tilde{\beta}_{\nu, 2m}^{(4)}.
\end{equation}
Apart from the sign, these equations are formally the same as those in
(\ref{4.8}) and (\ref{4.8a}). The stated formula thus follows as in
the derivation of (\ref{4.it}).

\vspace{.2cm}
Since the first sum is evaluated by the Christoffel-Darboux formula
(\ref{3.cd}), it remains to evaluate the second sum in
(\ref{4.it1}). This in turn requires the value of the
$\beta_{nk}^{(4)}$. Now, according to (\ref{3.gc}) with
$\tilde{\beta}_{2p+1,2p}^{(4)} = 0$, in the classical cases
the only non-zero value of
$\tilde{\beta}_{nk}^{(4)}$ for $n > k$ is 
$$
\tilde{\beta}_{2n, 2n-2}^{(4)} = - {c_{2n -1} \over c_{2n}}
$$
where $c_n$ is specified by (\ref{3.gam}) and (\ref{3.hlj}).
Hence in the classical cases
\begin{eqnarray}\label{4.sum3}
\tilde{S}_4(x,y) & = & {1 \over 2} e^{-({V}(y) - \tilde{V}(y))} 
e^{V(x) - \tilde{V}(x)}
{S}_2(x,y) \Big |_{N \mapsto 2N} \nonumber
\\&& +
{1 \over 2} e^{-\tilde{V}(x) + \tilde{V}(y) -
2 V(y) } {c_{2N-1} \over c_{2N}} {p_{2N}(y) \over (p_{2N}, p_{2N})_2}
\tilde{q}_{2N-2}^{(4)}(x).
\end{eqnarray}
We note that reference to the skew-orthogonal polynomial
$\tilde{q}_{2N-2}^{(4)}(x)$ can eliminated by noting from
(\ref{4.q4a}), and the fact that the only non-zero value of
$\tilde{\beta}_{\nu, 2m+1}^{(4)}$ is $\tilde{\beta}_{2m+1, 2m+1}^{(4)}
= 1$, that
\begin{equation}
e^{-\tilde{V}(x)} \tilde{q}_{2N-2}(x) =
-{\tilde{r}_{N-1}^{(4)} \over (p_{2N-1}, p_{2N-1})_2}
\int_{x}^\infty dt \, e^{-2V(t) + \tilde{V}(t)}
p_{2N-1}(t).
\end{equation}
Using (\ref{3.gc}) to eliminate $\tilde{r}_{N-1}^{(4)}$, and recalling
(\ref{3.gam}), from this
formula we thus have
\begin{eqnarray}\label{4.sum4}
\tilde{S}_4(x,y) & = & {1 \over 2} e^{-({V}(y) - \tilde{V}(y))}
e^{V(x) - \tilde{V}(x)}
{S}_2(x,y) \Big |_{N \mapsto 2N} \\ && -
{1 \over 2} e^{\tilde{V}(y) -
2 V(y) } \gamma_{2N-1} p_{2N}(y) 
\int_{x}^\infty dt \, e^{-2V(t) + \tilde{V}(t)}
p_{2N-1}(t).
\end{eqnarray}

In the Hermite case (\ref{4.sum4}) gives immediate agreement with the
formula in (\ref{4.tim}). However in the Laguerre case (\ref{4.sum4})
has a different structure to the formula in (\ref{4.tim}).
The manipulations necessary to show that the two formulas do indeed agree are
undertaken in the Appendix.

\section{Summary}
Classical skew orthogonal polynomials occur in the calculation of the
$n$-point distribution function $\rho_{(n)}$ associated with the
eigenvalue p.d.f.~(\ref{1.3}) with $w_\beta(x)$ a classical weight
function and $\beta = 1$ and 4. In particular they occur in a certain
sum, denoted $S_\beta(x,y)$, which determines $\rho_{(n)}$. This sum
is specified in (\ref{2.8a}) for $\beta = 1$ and $N$ even, in (\ref{2.kos})
for $\beta = 1$ and $N$ odd and in (\ref{2.10a}) for $\beta = 4$. The main
achievement of this paper has been the closed form evaluation of
$S_\beta(x,y)$ in terms of particular classical orthogonal polynomials
naturally related to the classical skew orthogonal polynomials. A self
contained presentation of the relationship between the classical
orthogonal and skew orthogonal polynomials is undertaken in Section 3.
The summation formulas
for $\tilde{S}_\beta(x,y)$ (the tilde on $\tilde{S}_\beta$ denotes that the
weight functions are modified according to (\ref{4.R1}) and
(\ref{4.R2})), which apply equally as well to all the classical
cases  are given by (\ref{4.sum1}) for $\beta = 1$ and $N$ even,
(\ref{4.sum2}) for $\beta = 1$ and $N$ odd and (\ref{4.sum4}) for
$\beta = 4$.

\subsection*{Acknowledgements}
M.A.~was supported by NSF grant \# DMS-98-4-50790, P.vM.~by
NSF grant \# DMS-98-4-50790, a FNRS and a Francqui Foundation grant,
while P.J.F.~and T.Nagao received support from the ARC. M.A., P.J.F.~and
P.vM.~are appreciative of the organisers of the MSRI Semester on
Random Matrices for facilitating this collaboration.

\section*{Appendix}
\setcounter{equation}{0}
\renewcommand{\theequation}{A.\arabic{equation}}
Here the summation formulas of (\ref{4.tim}) for $S_\beta(x,y)$ in the
Laguerre case will be shown to give agreement with expressions implied
by the general formulas (\ref{4.sum1}) with the substitutions
(\ref{3.hlj}) and (\ref{3.33'}). Consider first the case $\beta = 1$.
Because $\tilde{S}_1(x,y)$ is defined with the replacement
(\ref{4.R1}) and $f(x) = x$ in the Laguerre case, we see that we are required
to show
\begin{equation}\label{a.1}
S_1(x,y) = \tilde{S}_1(x,y) \Big |_{a \mapsto a + 1}.
\end{equation}
We see from (\ref{4.sum1}) and (\ref{3.33'}) that the r.h.s.~is presented as
a series in the linearly independent set of functions
$\{y^{a/2} e^{-y/2} L_k^{a+1}(y)\}$. On the other hand, the l.h.s.~can
also be expressed in terms of this basis by using the first identity
in (\ref{3.6'}) (with $a \mapsto a + 1$). Thus we must show that the
coefficients of each $y^{a/2} e^{-y/2}L_k^{a+1}(y)$ 
(which are functions of $x$) agree. The largest value of $k$ which occurs is
$k=N-1$. On the l.h.s.~the coefficient is then
\begin{equation}\label{a.1a}
x^{a/2} e^{-x/2} {(N-1)! \over \Gamma (a + N)} L_{N-1}^a(x) -
{N! \over 4 \Gamma ( a + N)} \int_0^\infty {\rm sgn} (x-u)
u^{a/2 - 1} e^{-u/2} L_N^{a-1}(u) \, du
\end{equation}
while on the r.h.s.~the coefficient is given by
\begin{equation}\label{a.1b}
- {(N-1)! \over 4 \Gamma (a + N)} \int_0^\infty {\rm sgn}(x-u)
L_{N-2}^{a+1}(u) u^{a/2} e^{-u/2} \, du.
\end{equation}
Now, by making use of the formula \cite{NF95}
\begin{equation}\label{a.nf}
\int_0^\infty x^{\alpha / 2} e^{-x/2} L_n^{\alpha + 1}(x) \, dx
= \left \{ \begin{array}{ll}{\Gamma ((n+3)/2)) \Gamma(\alpha + n + 2) \over
2^{\alpha / 2 - 1} \Gamma (n + 2) \Gamma ((n+\alpha +3)/2))}, & 
n  \: {\rm even}, \\[.2cm]
0, & n  \: {\rm odd} \end{array} \right.
\end{equation}
we can check that both (\ref{a.1a}) and (\ref{a.1b}) agree in the limit
$x \to \infty$. It thus suffices to show that both expressions have the
same derivative. This we can do by making use of the identities
(\ref{3.6'}) as well as the further identity
\begin{equation}\label{a.F}
xL_{N-1}^{a+1}(x) = - N L_N^{a+1}(x) + (2N + a) L_{N-1}^{a+1}(x) +
- (N + a) L_{N-2}^{a+1}(x).
\end{equation}

The coefficient of $y^{a/2} e^{-y/2}L_k^{a+1}(y)$ for $k < N$ 
on the l.h.s.~of (\ref{a.1}) is given by
\begin{equation}\label{a.2a}
\bigg ( {k! \over \Gamma (a + 1 + k)} L_k^a(x) -
{(k+1)! \over \Gamma (a + 2 + k)} L_{k+1}^a(x) \bigg )
x^{a/2} e^{-x/2},
\end{equation}
while on the r.h.s.~it is given by
\begin{equation}\label{a.2b}
\bigg ( {k! \over \Gamma (a + 2 + k)} x L_k^{a+1}(x) \bigg )
x^{a/2} e^{-x/2}.
\end{equation}
Use of (\ref{a.F}) and the first identity in (\ref{3.6'}) shows that
these expressions agree.

A similar procedure can be adopted to show that the formula in (\ref{4.tim})
for $S_4(x,y)$ in the Laguerre case agrees with that implied by
(\ref{4.sum4}). Specifically, the task is to show 
\begin{equation}\label{a.g}
S_4(x,y) = \tilde{S}_4(x,y) \Big |_{a \mapsto a - 1}.
\end{equation}
This we do by comparing coefficients of
$y^{a/2 - 1} e^{-y/2} L_k^{a-1}(y)$, $k=0,\dots,2N$. In the case
$k = 2N$ on the l.h.s.~use of (\ref{a.F}) shows this coefficient is equal to
$$
{1 \over 2} x^{a/2} e^{-x/2} {(2N-1)!)^2 L_{2N-1}^a(x) \over \Gamma (2N)
\Gamma (a + 2N) } (-2N) +
{(2N)! \over 4 \Gamma (2N + a)}
\int_0^x t^{a/2} e^{-t/2} \Big (
{d \over dt} L_{2N}^a(t) \Big ) \, dt
$$
while on the r.h.s. we read off from (\ref{4.sum4}), (\ref{3.hlj}) and 
(\ref{3.33'}) that the same coefficient is 
$$
{(2N)! \over 4 \Gamma (a + 2N - 1)} \int_x^\infty
t^{a/2 - 1} e^{-t/2} L_{2N-1}^{a-1}(t) \, dt.
$$
One now checks that both the expressions vanish as $x \to 0$
(for the latter this requires (\ref{a.nf})), and then verifies the
equality of the derivatives. Equating coefficients of
$y^{a/2 - 1} e^{-y/2} L_k^a(y)$ for $k < 2N$ gives 
an identity equivalent to the first formula in (\ref{3.6'}), thus verifying
their equality.

\end{document}